\documentclass[
    ,final            
  ]
 {aipproc} 
\layoutstyle{6x9}
\usepackage{color}

\usepackage{epsfig}
 \usepackage{rotating}
 \usepackage{amsmath,amssymb}
 \usepackage{graphicx}
\usepackage{ifthen,array}
 \usepackage{setspace}
 \usepackage{fancyhdr}
 \usepackage{moreverb}
 \usepackage{rotating}
\usepackage{amsfonts}
\usepackage{bbm}

\newcommand{\Sol}  {\textsc{sol}}

\newcommand{\Atm}  {\textsc{atm}}

\newcommand{\Dms}  {\Delta m^2_{21}}
\newcommand{\Dma}  {\Delta m^2_{31}}

\def\e6{$E(6)$}
\def\10{$SO(10)$}
\def\21{$SU(2) \otimes U(1) $}

\def\422{$SU(4) \otimes SU(2) \otimes SU(2)$}
\def\321{$SU(3) \otimes SU(2) \otimes U(1)$}
\def\lsim{\raise0.3ex\hbox{$\;<$\kern-0.75em\raise-1.1ex\hbox{$\sim\;$}}}
\def\gsim{\raise0.3ex\hbox{$\;>$\kern-0.75em\raise-1.1ex\hbox{$\sim\;$}}}
\def\lfv{lepton flavour violation }

\def\meff{\langle m_{\nu} \rangle}
\newcommand{\ed}{\end{document}}
\DeclareMathAlphabet{\mathsc}{OT1}{cmr}{m}{sc}

\def \znbb {$0\nu\beta\beta$ }

\def\meff{\langle m_{\nu} \rangle}

\let\vev\VEV

\def\e6{$E(6)$}
\def\10{$SO(10)$}
\def\21{$SU(2) \otimes U(1) $}

\def\422{$SU(4) \otimes SU(2) \otimes SU(2)$ }
\def\321{$SU(3) \otimes SU(2) \otimes U(1)$ }

\renewcommand{\baselinestretch}{1.07}

\newcommand{\AddrAHEP}{%
 AHEP Group, Instituto de F\'{\i}sica Corpuscular,
  C.S.I.C. -- Universitat de Val{\`e}ncia \\
  Edificio de Institutos de Paterna, Apartado 22085,
  E--46071 Val{\`e}ncia, Spain\\}

\begin{document}
\newcommand{\Od}{{\cal O}}

\title{Neutrino mass in supersymmetry}

\author{J. W. F. Valle}{
address={\AddrAHEP}}

\begin{abstract}

  After summarizing neutrino oscillation results I discuss high and
  low-scale seesaw mechanisms, with or without supersymmetry, as well
  as recent attempts to understand the pattern of neutrino mixing from
  flavor symmetries.  I also mention the possibility of intrinsic
  supersymmetric neutrino masses in the context of broken R parity
  models, showing how this leads to clear tests at the LHC.

\end{abstract}
\maketitle


\section{neutrino oscillations}
\label{sec:neutr-oscill}

\begin{figure}[t]
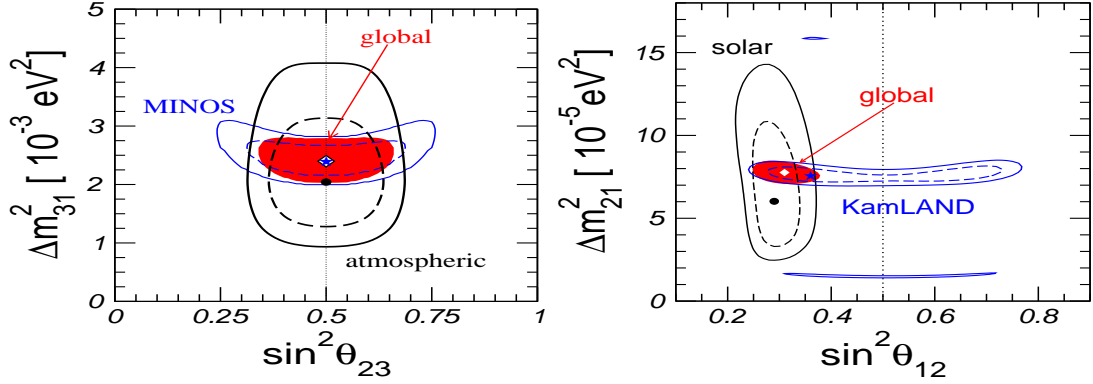
 \centering
\includegraphics[width=.48\linewidth,height=5cm]{atm-new.eps}
\includegraphics[width=.48\linewidth,height=5cm]{sol-new.eps}
\caption{\label{fig:global} %
  Current neutrino oscillation parameters, from
  Ref.~\cite{Schwetz:2008er}. }
\end{figure}

\begin{figure}[h]
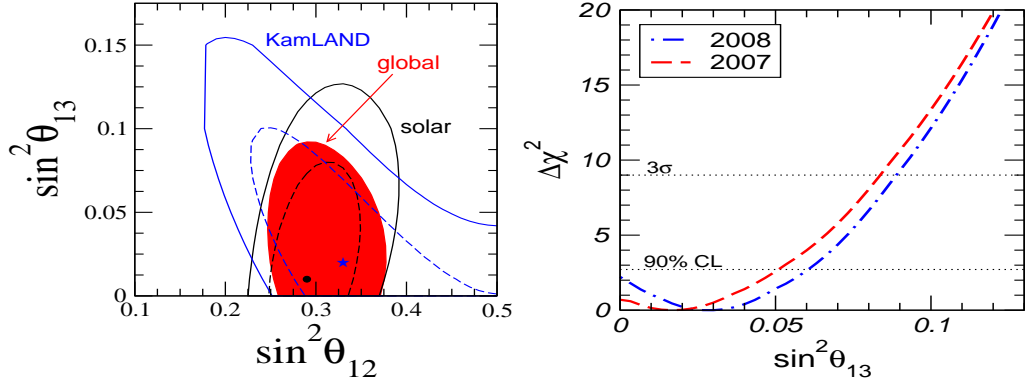
 \centering
\includegraphics[height=5cm,width=.45\linewidth]{s12-s13-tension.eps}
\includegraphics[height=5cm,width=.45\linewidth]{th13-sol-07vs08-lin.eps}
\caption{\label{fig:th13}%
Constraints on $\sin^2\theta_{13}$ from different parts of
  the global data given in Ref.~\cite{Schwetz:2008er}.}
\end{figure}

We now have uncontroversial evidence for neutrino flavor conversion
coming from ``celestial'' (solar and atmospheric) as well as
``laboratory'' studies with reactor and accelerator
neutrinos~\cite{Maltoni:2004ei,Schwetz:2008er}.
Oscillations constitute the only viable explanation of the data and
provide the first sign of physics beyond the Standard Model (SM). The
basic concept in terms of which to describe them is the lepton mixing
matrix, the leptonic analogue of the quark mixing matrix. In its
simplest $3\times 3$ unitary form it is given
as~\cite{schechter:1980gr}
\begin{equation}
  \label{eq:2227}
K =  \omega_{23} \omega_{13} \omega_{12}
\end{equation}
where each $\omega$ is characterized by an angle and a corresponding
CP phase. Present experiments are insensitive to CP violation, hence
we set all three phases to zero.
In this approximation oscillations depend on the three mixing angles
$\theta_{12}, \theta_{23}, \theta_{13}$ and on the two squared-mass
splittings $\Dms \equiv m^2_2 - m^2_1$ and $\Dma \equiv m^2_3 - m^2_1$
characterizing solar and atmospheric transitions.  To a good
approximation, one can set $\Dms = 0$ in the analysis of atmospheric
and accelerator data, and $\Dma$ to infinity in the analysis of solar
and reactor data.  The neutrino oscillation parameters obtained from a
global analysis of the world's neutrino oscillation data are
summarized in Figs.~\ref{fig:global} and \ref{fig:th13}. The left and
right panels in Fig.~\ref{fig:global} give the ``atmospheric'' and
``solar'' oscillation parameters, $\theta_{23}$ \& $\Dma$, and
$\theta_{12}$ \& $\Dms$, respectively. 
The dot, star and diamond indicate the best fit points of atmospheric
MINOS and global data, respectively. We minimize with respect to
$\Dms$, $\theta_{12}$ and $\theta_{13}$, including always all the
relevant data.
Similarly the ``solar'' oscillation parameters are obtained by
combining solar and reactor neutrino data. 
The dot, star and diamond indicate the best fit points of solar,
KamLAND and global data, respectively. We minimize with respect to
$\Dma$, $\theta_{23}$ and $\theta_{13}$, including always all relevant
data.
One sees that data from artificial and natural neutrino sources are
clearly complementary: reactor and accelerators give the best
determination of squared-mass-splittings, while solar and atmospheric
data mainly determine mixings.
Fig.~\ref{fig:th13} summarizes the information on the remaining angle
$\theta_{13}$, the right panel shows how current data slightly prefer
a nonzero value for $\theta_{13}$. Since this is currently not
significant, we prefer to interpret this as a weaker bound on
$\theta_{13}$~\footnote{Note: the bounds in Eq.~(\ref{eq:th13a}) are
  given for 1~dof, while the regions in Fig.~\ref{fig:th13} (left) are
  90\%~CL for 2~dof}:

\begin{equation}\label{eq:th13a}
  \sin^2\theta_{13} \le \left\lbrace \begin{array}{l@{\qquad}l}
      0.060~(0.089) & \text{(solar+KamLAND)} \\
      0.027~(0.058) & \text{(CHOOZ+atm+K2K+MINOS)} \\
      0.035~(0.056) & \text{(global data)}
    \end{array} \right.
\end{equation}
A possible experimental confirmation of a non-zero $\theta_{13}$ would
encourage the search for CP violation in upcoming neutrino oscillation
experiments~\cite{Bandyopadhyay:2007kx,Nunokawa:2007qh}.
Note that all CP violating observables, such as CP asymmetries, are
proportional to the small parameter $ \alpha \equiv \frac{\Delta
  m^2_{21}}{|\Delta m^2_{31}|}$ well-determined experimentally as $
\alpha = 0.032\,, \quad 0.027 \le \alpha \le 0.038 \quad (3\sigma) \,.
$

\section{On the origin of neutrino mass}
\label{sec:neutrino-mass}

\begin{figure}[!h] \centering
 \includegraphics[height=3cm,width=.4\linewidth]{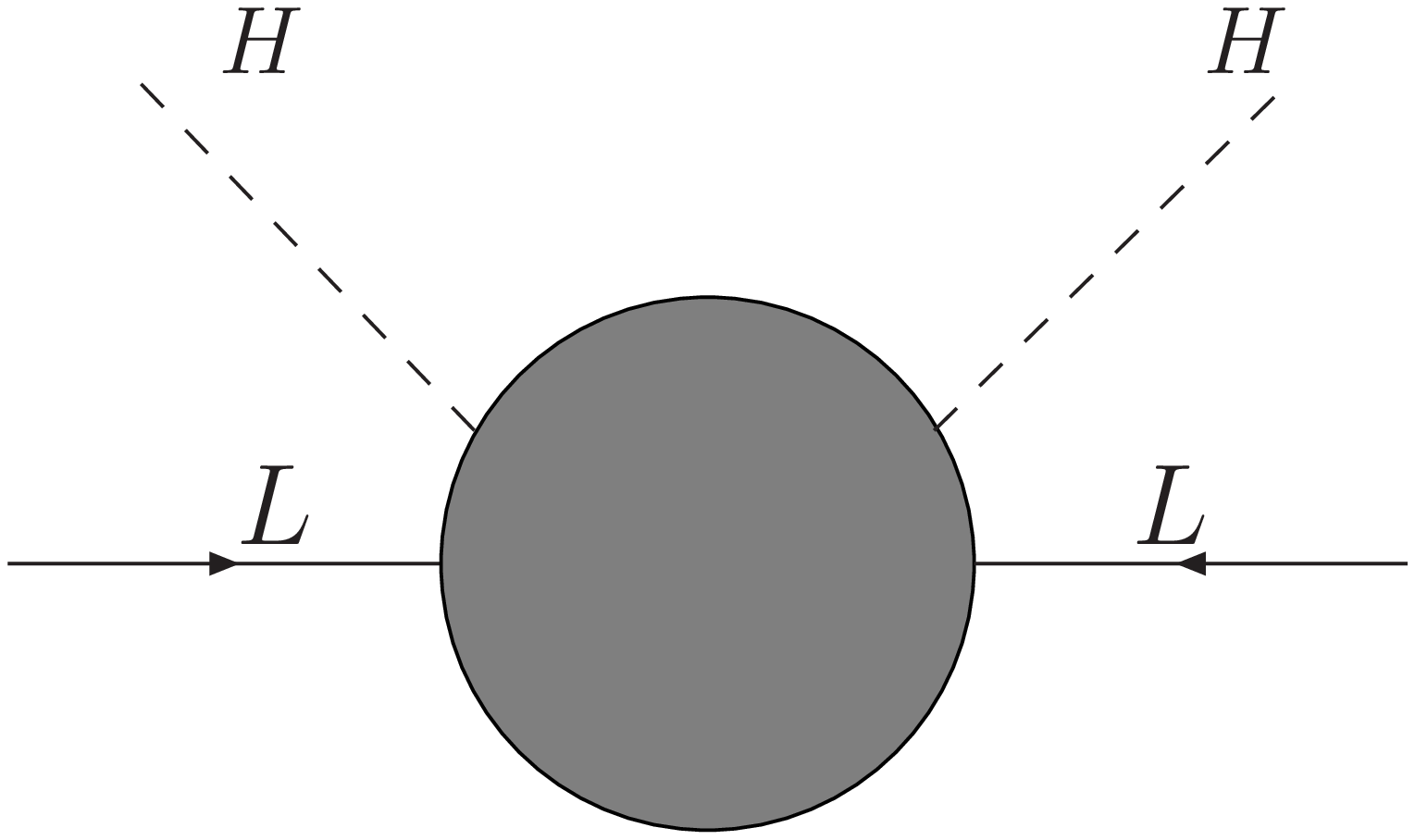}
\hskip 1cm
 \includegraphics[height=3cm,width=.4\linewidth]{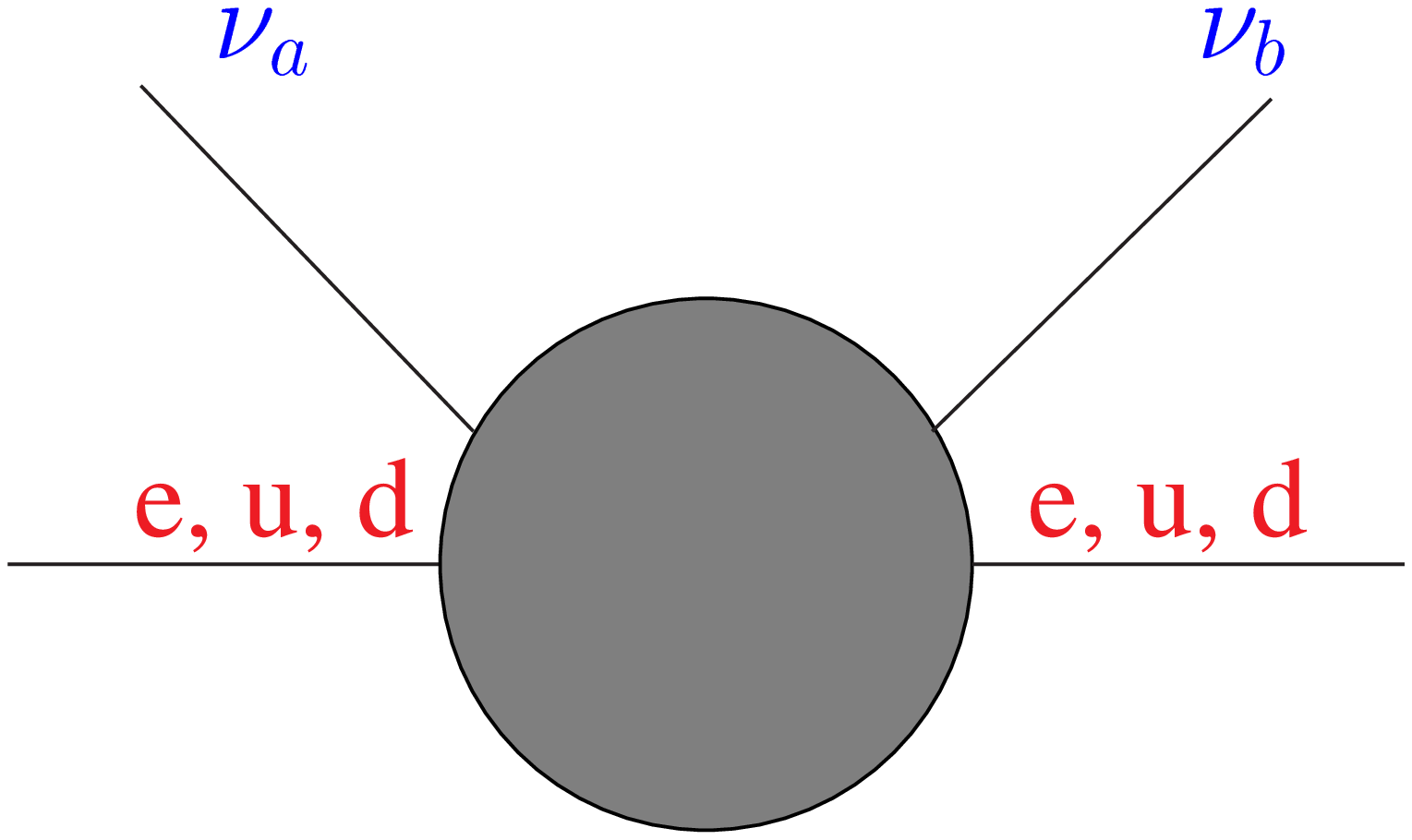}
 \caption{\label{fig:d-5-nsi} %
   Operators characterizing neutrino masses and non-standard neutrino
   interactions (NSI) arising, say, from the non-trivial structure of
   lepton mixing in seesaw-type schemes~\cite{schechter:1980gr}.}
\end{figure}

In spite of the great experimental progress summarized above,
pinning-down the ultimate origin of neutrino mass remains a challenge
for the next decades. To understand the pathways to neutrino masses it
is important to note that, being electrically neutral, neutrino masses
should, on general grounds, be of
Majorana-type~\cite{schechter:1980gr}.  Indeed, specific neutrino mass
generation mechanisms also differ from those of charged fermions in
the SM. The latter come in two chiral species and get mass linearly in
the electroweak symmetry breaking vacuum expectation value (vev)
$\vev{\Phi}$ of the Higgs scalar doublet.  In contrast, as shown in
the left panel in Fig.~\ref{fig:d-5-nsi}, neutrinos acquire mass from
an effective lepton number violating dimension-five operator $\lambda
L \Phi L \Phi$ (where $L$ denotes a lepton
doublet)~\cite{Weinberg:1980bf}.

A natural way to account for the smallness of neutrino masses,
irrespective of their specific origin, is that L-number is restored,
in the absence of L-violating operator(s).
Such may be naturally suppressed either by a high-scale $M_X$ in the
denominator or, alternatively, it may involve a low-mass-scale in the
numerator.
The big question is to identify which \emph{ mechanism} gives rise to
this operator, its associated mass \emph{ scale} and its \emph{flavor
  structure}.
Gravity is often argued to break global
symmetries~\cite{Coleman:1988tj,Kallosh:1995hi}, and could induce the
dimension-five operator, with $M_X$ identified to the Planck
scale. The resulting Majorana neutrino masses are too small, hence 
the need for physics beyond the SM~\cite{deGouvea:2000jp}.

The coefficient $\lambda$ could vanish due to symmetry, so that the
effective operator responsible for neutrino mass is of higher
dimension~\cite{Gogoladze:2008wz,Bonnet:2009ej}.
Alternatively it may be suppressed by small scales, Yukawa couplings
and/or loop-factors~\cite{Valle:2006vb}. To arrange our brief
discussion I consider three options: (i) tree level, (ii) radiative,
and (iii) hybrid mechanisms, all of which may have high- or low-scale
realizations.
If lepton-number symmetry is broken spontaneously there is either an
extra neutral gauge boson or a Nambu-Goldstone boson coupled to
neutrinos, depending on whether it is gauged or not.
It is easy to construct models based on either high- or low-scale
symmetry breaking, the former are more popular among theorists,
because they are closer to the idea of unification.

However the most basic and general description of the seesaw is in
terms of the SM \321 gauge structure~\cite{schechter:1980gr}. In such
a framework it is clear that the relevant scale can be large or small,
depending on model details. Hence there is a fair chance that the
origin of neutrino mass may be probed at accelerators like the LHC.

\vglue -0.5cm
\subsection{(i)  Minimal seesaw schemes }
\label{sec:i-minimal-seesaw}

The classic way to to generate Weinberg's dimension-5
operator~\cite{Weinberg:1980bf} is the exchange of heavy fermion
states with masses close to the ``unification'' scale.  \vglue .3cm
\begin{figure}[h] 
\centering
 \includegraphics[scale=.3,width=.45\linewidth]{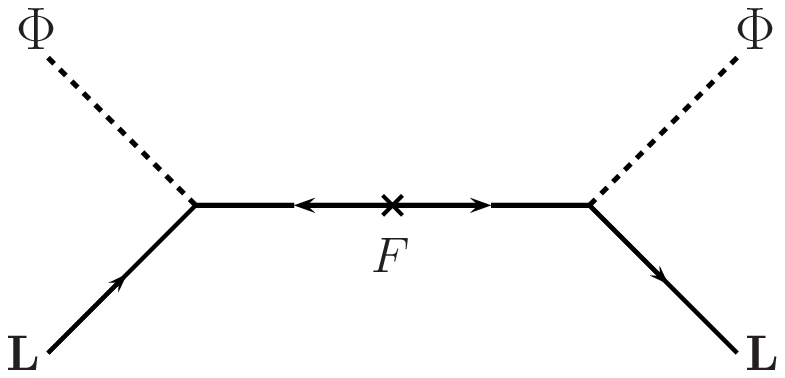} 
 \includegraphics[height=2.5cm,width=.45\linewidth]{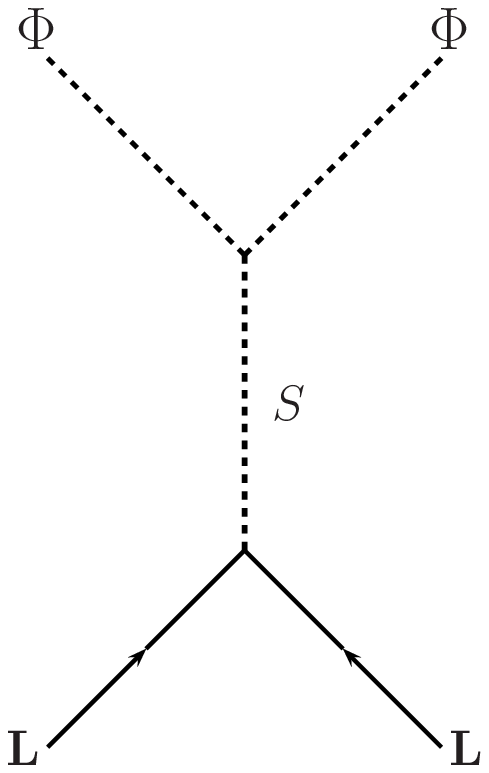}
 \caption{\label{fig:seesaw}
   Type-I~
   and III~(left) and 
   Type-II~(right) 
   realizations of the seesaw mechanism.}
 \end{figure}
 Depending on whether these are \321 singlets or triplets the
 mechanism is called
 \textbf{type-I}~\cite{schechter:1980gr}~\cite{Minkowski:1977sc,gell-mann:1980vs,yanagida:1979,mohapatra:1980ia},
 or \textbf{type-III} seesaw~\cite{foot:1998iw}, respectively. As seen
 in the right panel in Fig.~\ref{fig:seesaw}, the seesaw may also be
 induced by the exchange of heavy triplet scalars, now called
 \textbf{type-II}
 seesaw~\cite{schechter:1980gr}~\cite{schechter:1982cv,Lazarides:1980nt},
 a convention opposite to the one used originally in
 \cite{schechter:1980gr}.
 The ``complete seesaw'' was thoroughly studied
 in~\cite{schechter:1980gr} and its perturbative diagonalization was
 given Ref.~\cite{schechter:1982cv} in a general form that may be
 adapted to different models. The hierarchy of vevs required to
 account the small neutrino masses $v_3 \ll v_2 \ll v_1$ in such
 seesaw was studied in detail in Ref.~\cite{schechter:1982cv}. %

\subsection{(ii)  ``Non-minimal'' seesaw schemes}
\label{sec:ii-non-missionaire}

The seesaw may be implemented in the type I, type II or type III
manner, with different gauge groups and multiplet contents, with
gauged or ungauged B-L, broken explicitly or spontaneously, at a high
or at a low energy scale, with or without supersymmetry.
There are so many ways to seesaw, that a full taxonomy describing all
variants will probably never be written, as nature may be more
imaginative than physicists.
Since any extended symmetry model must ultimately break to the SM,
what is phenomenologically relevant is the seesaw description at the
\321 level~\cite{schechter:1980gr}. Such low-energy description is
specially relevant in accurately describing low-scale variants of the
seesaw mechanism, whose interest has now been revived with the coming
of the LHC.
 An attractive class of such schemes employs, in addition to the
 left-handed SM neutrinos $\nu_L$, two \321 singlets $\nu^c, \,
 S$~\cite{mohapatra:1986bd} (for other extended seesaw schemes see,
 e.g.~\cite{Wyler:1983dd,GonzalezGarcia:1988rw,Akhmedov:1995vm,Barr:2005ss}).
 The basic lepton-number-violating parameter is
 small~\cite{Hirsch:2009mx,Ibanez:2009du} and may be calculable due to
 supersymmetric renormalization group evolution
 effects~\cite{Bazzocchi:2009kc}.
 One may implement such schemes in the \10
 framework~\cite{Malinsky:2005bi,Hirsch:2006ft}, leaving, in addition
 a light $Z^\prime$ to be probed at the LHC~\cite{valle:1987sq}.
\begin{figure}[b]
\vspace{9pt}
\includegraphics[width=6cm,height=45mm]{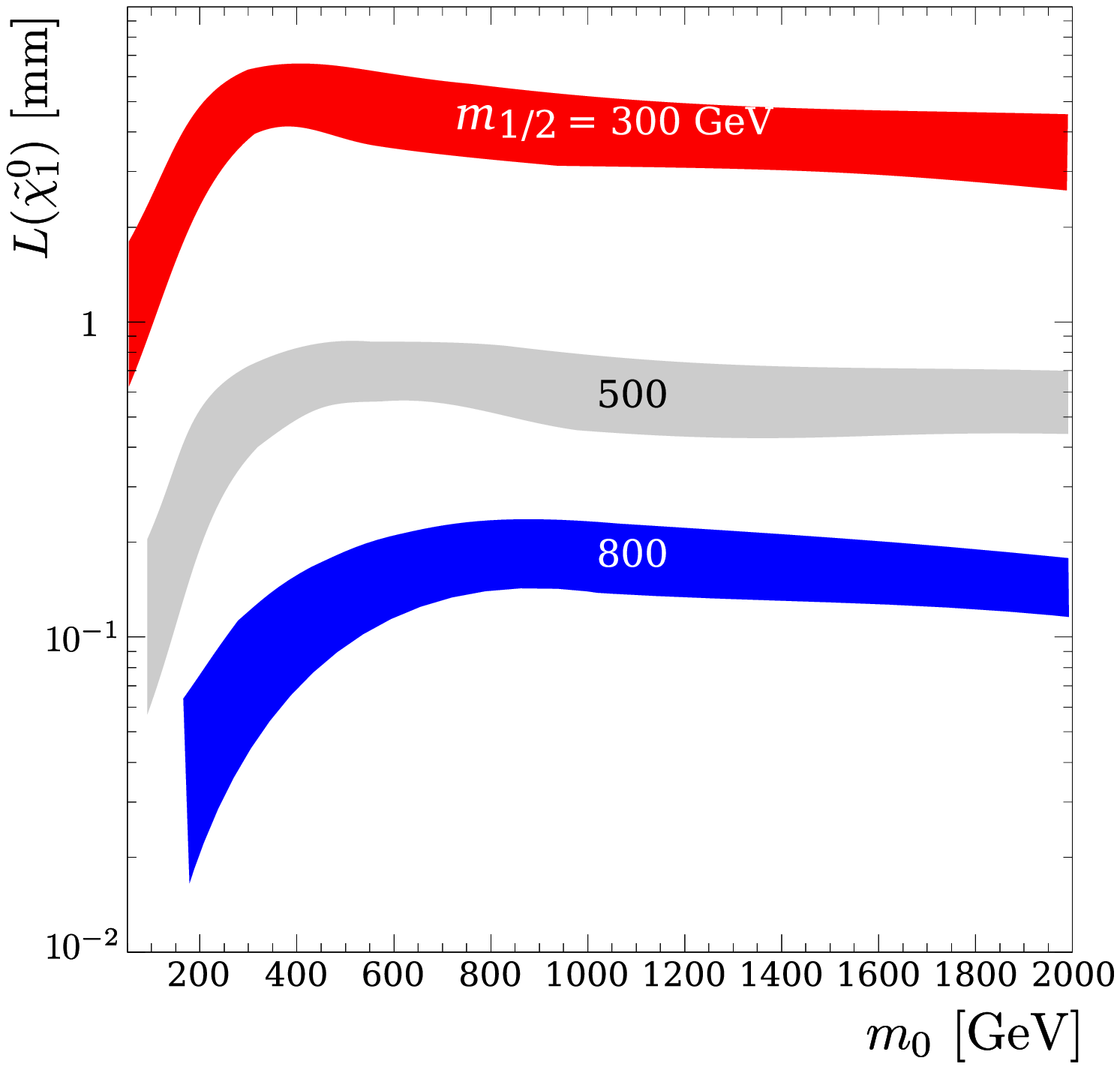}
\includegraphics[width=60mm,height=48mm]{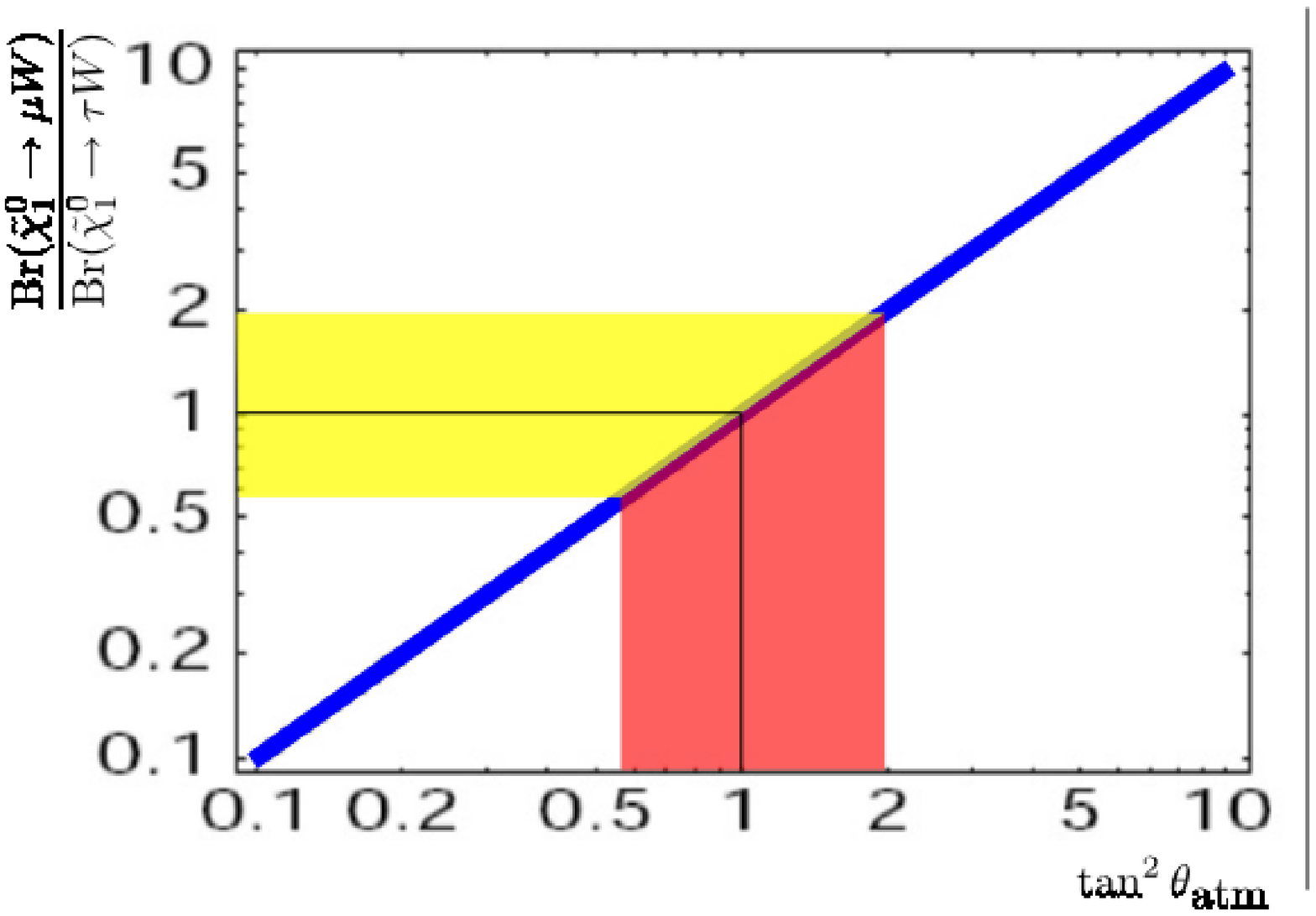}
\caption{Left: $\tilde\chi_1^0$ decay length versus $m_0$ for
  $A_0=-100$ GeV, $\tan\beta=10$, $\mu > 0$, and several values of
  $m_{1/2}$.  The widths of the three shaded bands around
  $m_{1/2}=300,~500,~800$ GeV correspond to the variation of the BRpV
  parameters in such a way that the neutrino masses and mixing angles
  fit the required values within $3\sigma$. Right: Ratio of branching
  ratios, Br$(\chi^0_1\to \mu q'{\bar q})$ over Br$(\chi^0_1\to \tau
  q'{\bar q})$ as a function of the atmospheric angle in bilinear R
  parity violation~\cite{Porod:2000hv}.}
\label{fig:ntrl}
\end{figure}

\vglue -0.5cm
\subsection{(iii) Radiative schemes}
\label{sec:ii-radiative-schemes}

Neutrino masses may be absent at tree level and
calculable~\cite{zee:1980ai,babu:1988ki}, with no need for a large
scale.
In this case the coefficient $\lambda$ is suppressed by small
loop-factors, by Yukawa couplings and possibly by a small scale
parameter characterizing the breaking of lepton number, leading to
naturally small neutrino masses.
Like low-scale seesaw schemes (see above) radiative models open the
door to phenomenology associated with the new states required to
provide the neutrino mass and which could be searched for, e.~g., at
the LHC.

\vglue -0.5cm

\subsection{(iv) R parity violation}
\label{sec:vvv rpv}

It could well be that the origin of neutrino masses is intrinsically
supersymmetric. This is the case in models with R parity
violation~\cite{hall:1984id,Ross:1984yg,Ellis:1984gi}, in which lepton
number is broken together with the so-called R parity.
This may happen spontaneously, driven by a nonzero vev of an
\321 singlet
sneutrino~\cite{Masiero:1990uj,romao:1992vu,romao:1997xf}, leading to
an effective model with bilinear violation of R
parity~\cite{Diaz:1998xc,Hirsch:2004he}. 
This provides the minimal way to break R parity and add neutrino
masses to the MSSM~\cite{Hirsch:2004he}.  The neutrino spectrum is
hybrid, with one scale (typically the atmospheric) generated at tree
level by neutralino-exchange \emph{weak-scale seesaw}, and the other
scale (solar) induced by \emph{calculable} one-loop
corrections~\cite{Hirsch:2000ef}.

Unprotected by any symmetry, the lightest supersymmetric particle
(LSP) will decay.  Given the scale of neutrino mass indicated by
experiment these decays will happen inside typical detectors at the
Tevatron or the LHC~\cite{Hirsch:2000ef,Diaz:2003as} but with a decay
path that can be experimentally resolved, leading to a so-called
displaced vertex~\cite{deCampos:2005ri,deCampos:2007bn} (left panel in
Fig.~\ref{fig:ntrl}).
More strikingly, its decay properties correlate with the neutrino
mixing angles. Indeed, as seen in the right panel in
Fig.~\ref{fig:ntrl} the LSP decay pattern is predicted by the
low-energy measurement of the atmospheric
angle~\cite{Porod:2000hv,romao:1999up,mukhopadhyaya:1998xj}.
Such a prediction will be tested at the LHC, and will potentially
allow a high-energy redetermination of $\theta_{23}$.
Similar correlations hold in variant models which have other
supersymmetric states as LSP~\cite{Hirsch:2003fe}.

\vglue -0.5cm
\section{Flavor symmetries}
\label{sec:Flavor}

As seen above current neutrino oscillation data indicate solar and
atmospheric mixing angles which are unexpectedly large when compared
with quark mixing angles. This challenges our attempts to explain the
flavor problem in unified schemes where quarks and leptons are
related. 
It has been noted that the neutrino mixing angles are approximately
given by~\cite{Harrison:2002er},
\begin{align}
\label{eq:hps}
\tan^2\theta_{\Atm}&=\tan^2\theta_{23}^0=1\\ \nonumber
\sin^2\theta_{\textrm{Chooz}}&=\sin^2\theta_{13}^0=0\\
\tan^2\theta_{\Sol}&=\tan^2\theta_{12}^0=0.5 .\nonumber
\end{align}
There have been many schemes suggested in the literature in order to
reproduce the full tri-bi-maximal pattern, or at least to predict
maximal atmospheric mixing using various discrete flavor symmetry
groups containing mu-tau symmetry,
e.~g.~\cite{babu:2002dz,Hirsch:2003dr,Harrison:2002et,Grimus:2003yn,
  Altarelli:2005yp,Mondragon:2007af,Bazzocchi:2009da,Altarelli:2009gn,Grimus:2009mm,Joshipura:2009tg}.
One expects the flavor symmetry to be valid at high energy scales.
Deviations from tri-bi-maximal ansatz~\cite{King:2009qt} may be
calculable by renormalization group
evolution~\cite{Antusch:2003kp,Plentinger:2005kx,Hirsch:2006je}.

A specially simple ansatz is that, as a result of a given flavor
symmetry such as A4~\cite{babu:2002dz,Hirsch:2003dr}, neutrino masses
unify at high energies $M_X$~\cite{chankowski:2000fp}, the same way as
gauge couplings unify at high energies due to
supersymmetry~\cite{amaldi:1991cn}. Such quasi-degenerate neutrino
scheme predicts maximal atmospheric angle and vanishing $\theta_{13}$,
$$\theta_{23}=\pi/4~~~\rm{and}~~~\theta_{13}=0\:,$$ 
leaving the solar angle $\theta_{12}$ unpredicted, but
Cabibbo-unsuppressed,
$$\theta_{12}={\cal O}(1).$$ 
If CP is violated $\theta_{13}$ becomes arbitrary and the Dirac phase
is maximal~\cite{Grimus:2003yn}.  One can show that lepton and slepton
mixings are related and that at least one slepton lies below 200 GeV,
within reach of the LHC. The absolute Majorana neutrino mass scale
$m_0 \gsim 0.3$ eV ensures that the model will be probed by future
cosmological tests and $\beta\beta_{0\nu}$ searches.  Rates for lepton
flavour violating processes $l_j \to l_i + \gamma$ typically lie in
the range of sensitivity of coming experiments, with BR$(\mu \to e
\gamma) \gsim 10^{-15}$ and BR$(\tau \to \mu \gamma) > 10^{-9}$.

\vglue -0.5cm
\section{Lepton flavor violation (LFV)}
\label{sec:lfv}

Flavor is violated in neutrino
propagation~\cite{Maltoni:2004ei,Schwetz:2008er}. It is therefore
natural to expect that, at some level, it will also show up as
transitions involving the charged leptons, since these sit in the same
electroweak doublet.  Two basic mechanisms are: (i) neutral heavy
lepton
exchange~\cite{Bernabeu:1987gr,gonzalez-garcia:1992be,Ilakovac:1994kj}
and (ii)
supersymmetry~\cite{borzumati:1986qx,casas:2001sr,Antusch:2006vw}.
Both exist in supersymmetric seesaw-type schemes of neutrino mass,
the interplay of both types of contributions depends on the seesaw
scale and has been analysed in ~\cite{Deppisch:2004fa}.
Barring fine-tunings, high-scale seesaw models require supersymmetry
in order to have sizeable LFV rates.  Moreover, supersymmetry brings
in the possibility of direct \lfv in the production of supersymmetric
particles. This will provide the most direct way to probe LFV at the
LHC in high-scale seesaw models, as seen in Fig.~\ref{fig:ntrl}, from
Ref.~~\cite{Esteves:2009vg}.
\begin{figure}[!htb]
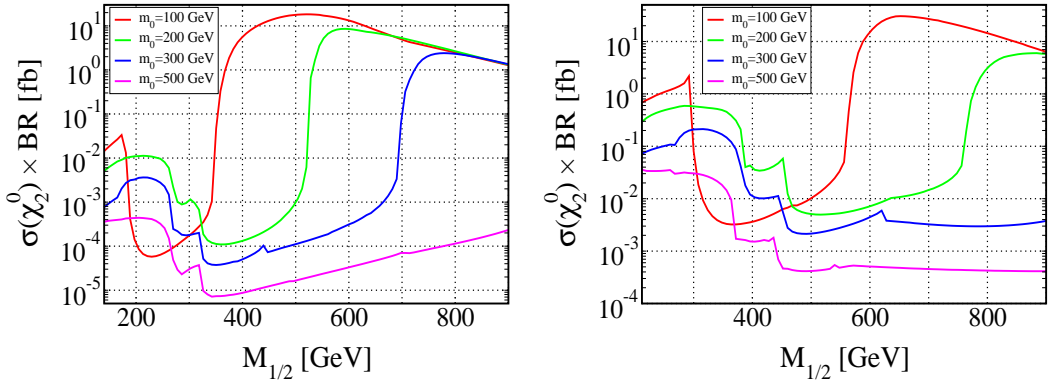

  \centering
\begin{tabular}{cc}
  \includegraphics[width=0.45\textwidth,height=5cm]{plot-sigmaLOfbBR-m12_-_4m0.eps} &
  \includegraphics[width=0.45\textwidth,height=5cm]{plot-sigmaLOfbBR-m12_-_4m0_-_II.eps}
\end{tabular}
\caption{LFV rate for $\mu$-$\tau$ lepton pair production from
  $\chi^0_2$ decays versus $M_{1/2}$ for the indicated $m_0$ values,
  assuming minimal supergravity parameters: $\mu>0$, $\tan\beta=10$
  and $A_0=0$ GeV, for type-I (left) and for type-II seesaw
  (right). Here $\lambda_1=0.02$ and $\lambda_2=0.5$ are Type-II
  seesaw parameters, and we imposed the contraint Br($\mu\to e
  +\gamma) \le 1.2\cdot 10^{-11}$.}
  \label{fig:ProdXBR}
\end{figure}

\begin{figure}[!h]
  \centering
\includegraphics[width=7cm,height=5cm]{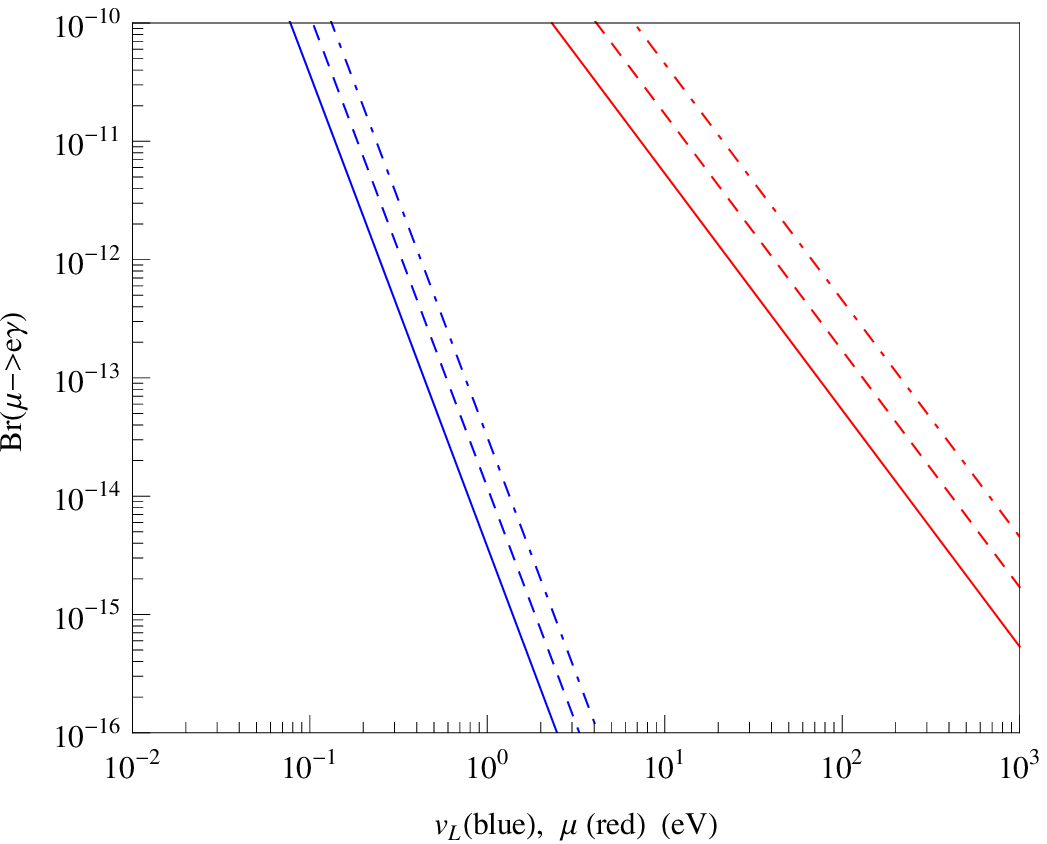}
\includegraphics[width=7cm,height=5cm]{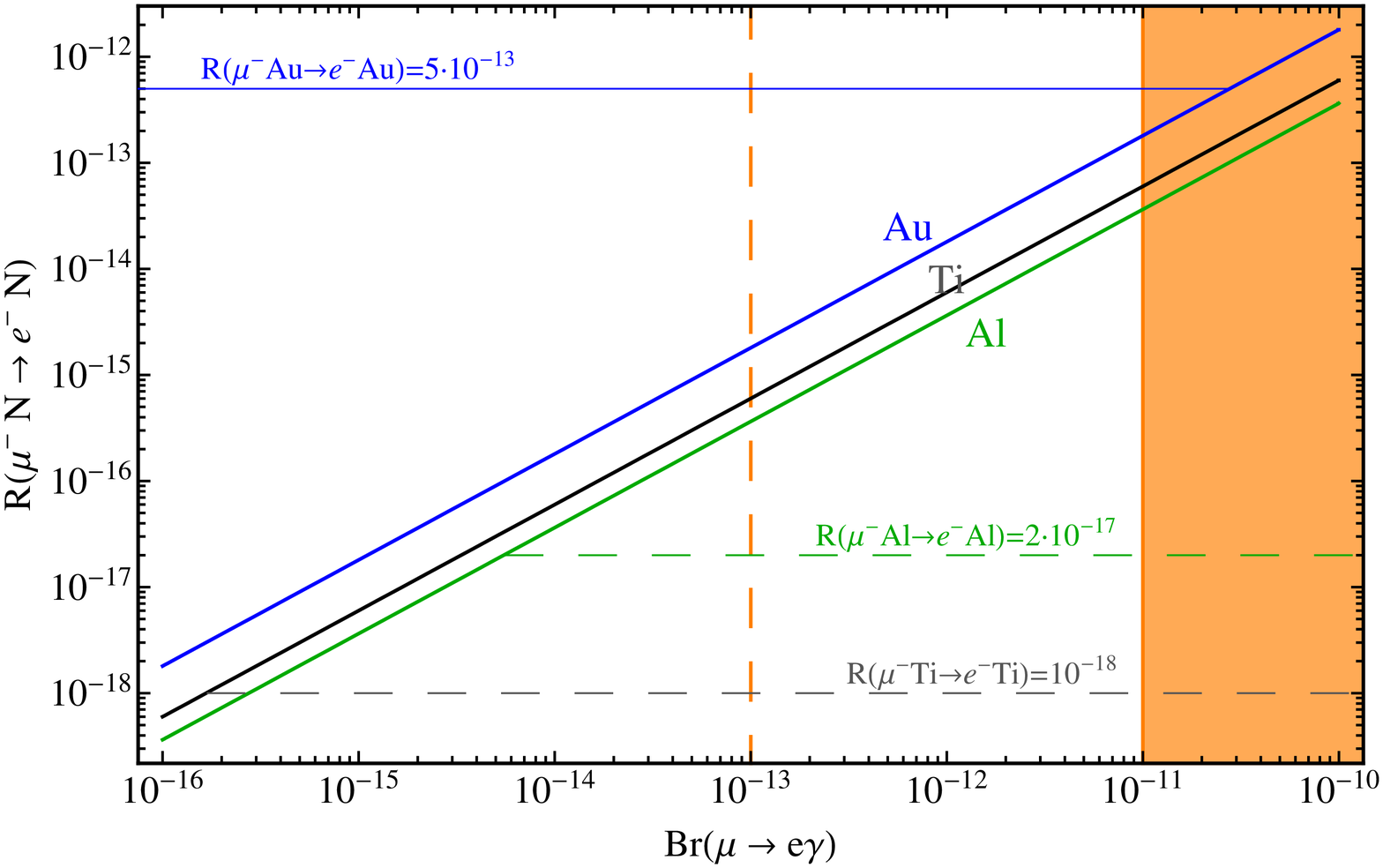}
\caption{Left: $Br(\mu\to e \gamma)$ versus the lepton number
  violation scale for two low-scale seesaw models: the inverse seesaw
  in top (red color), the linear seesaw in bottom (blue color). In
  both cases, $M$ is fixed as $M=100 \, GeV$ (continous line),
  $M=200\, GeV$ (dashed line) and $M=1000\, GeV$ (dot-dashed
  line). The right panel shows typical correlation between mu-e
  convertion in nuclei and \(Br(\mu\to e\gamma)\).}
\label{fig:lfv-low-scale}
\end{figure}

In contrast, the sizeable admixture of right-handed neutrinos in the
charged current (rectangular nature of the lepton mixing
matrix~\cite{schechter:1980gr}) in low-scale seesaw schemes induces
potentially large LFV rates even in the absence of
supersymmetry~\cite{Bernabeu:1987gr}.  
Indeed, an important point to stress is that
LFV~\cite{Bernabeu:1987gr,gonzalez-garcia:1992be} and CP
violation~\cite{branco:1989bn,Rius:1989gk} can occur in the massless
neutrino limit, hence their attainable magnitude is unrestricted by
the smallness of neutrino masses.
In Fig.~\ref{fig:lfv-low-scale} we display \(Br(\mu\to e\gamma)\)
versus the small lepton number violating (LNV) parameters $\mu$ and
$v_L$ for two different low-scale seesaw models, the inverse and the
linear seesaw, respectively.  Clearly the LFV rates are sizeable in
both cases, the different slopes with respect to $\mu$ and $v_L$
follow from the fact that LNV occurs differently in the two models,
$\Delta L=2$ versus the $\Delta L=1$, respectively.
Similarly one can show~\cite{Deppisch:2005zm} that in low-scale seesaw
models the nuclear $\mu^--e^-$ conversion rates lie within planned
sensitivities of future experiments such as PRISM~\cite{Kuno:2000kd}.
Note that models with specific flavor symmetries, such as those in
\cite{Hirsch:2009mx,Ibanez:2009du} relate different LFV rates. To
conclude we mention that the some seesaw schemes, like
type-III~\cite{foot:1998iw} or inverse type-III~\cite{Ibanez:2009du},
may be directly probed at the LHC by directly producing the TeV RH
neutrinos at accelerators.

\vglue -0.5cm
\section{Lepton number violation}
\label{sec:lepton-number-lepton}

\begin{figure}[h]
  \centering
\includegraphics[width=5cm,height=4cm]{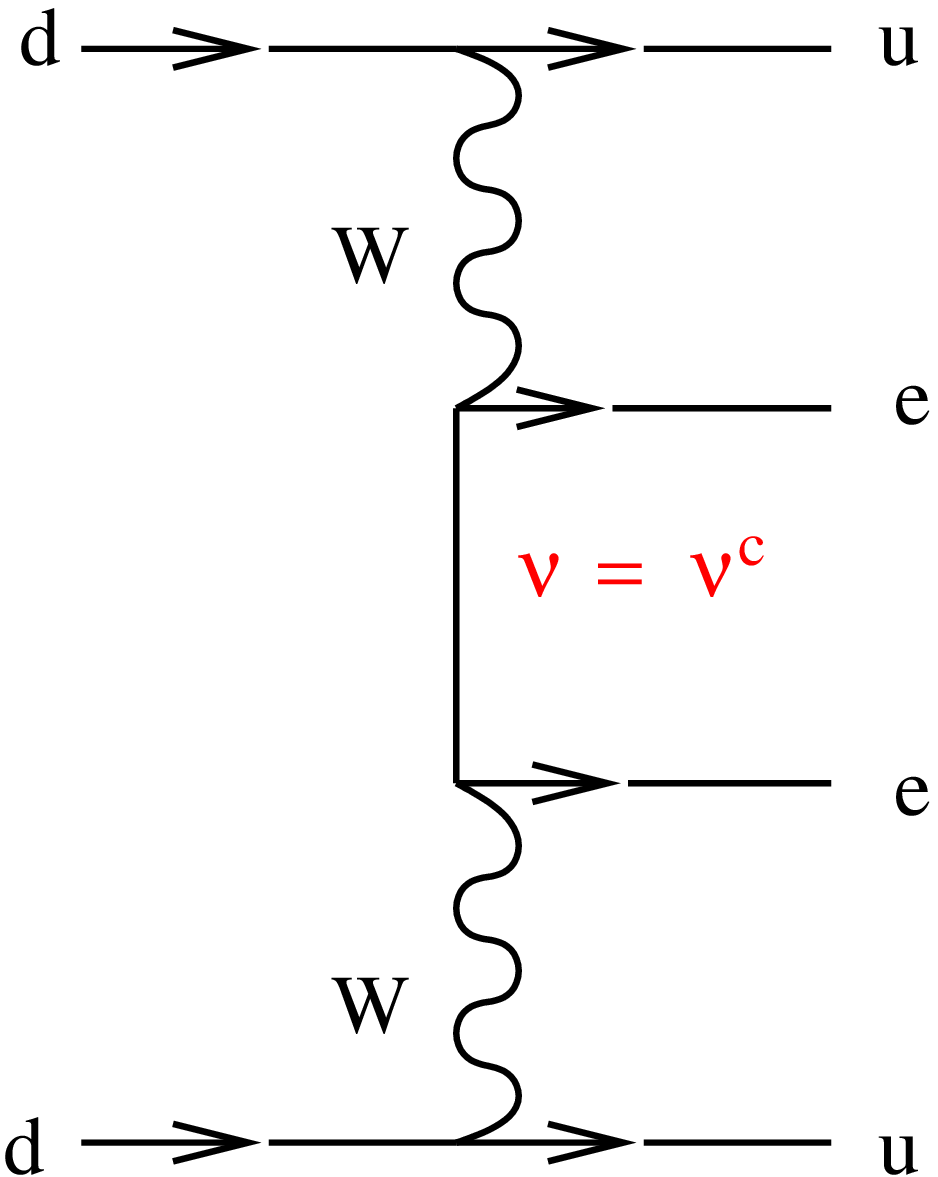} 
\hglue 1cm
\includegraphics[width=5cm,height=4cm]{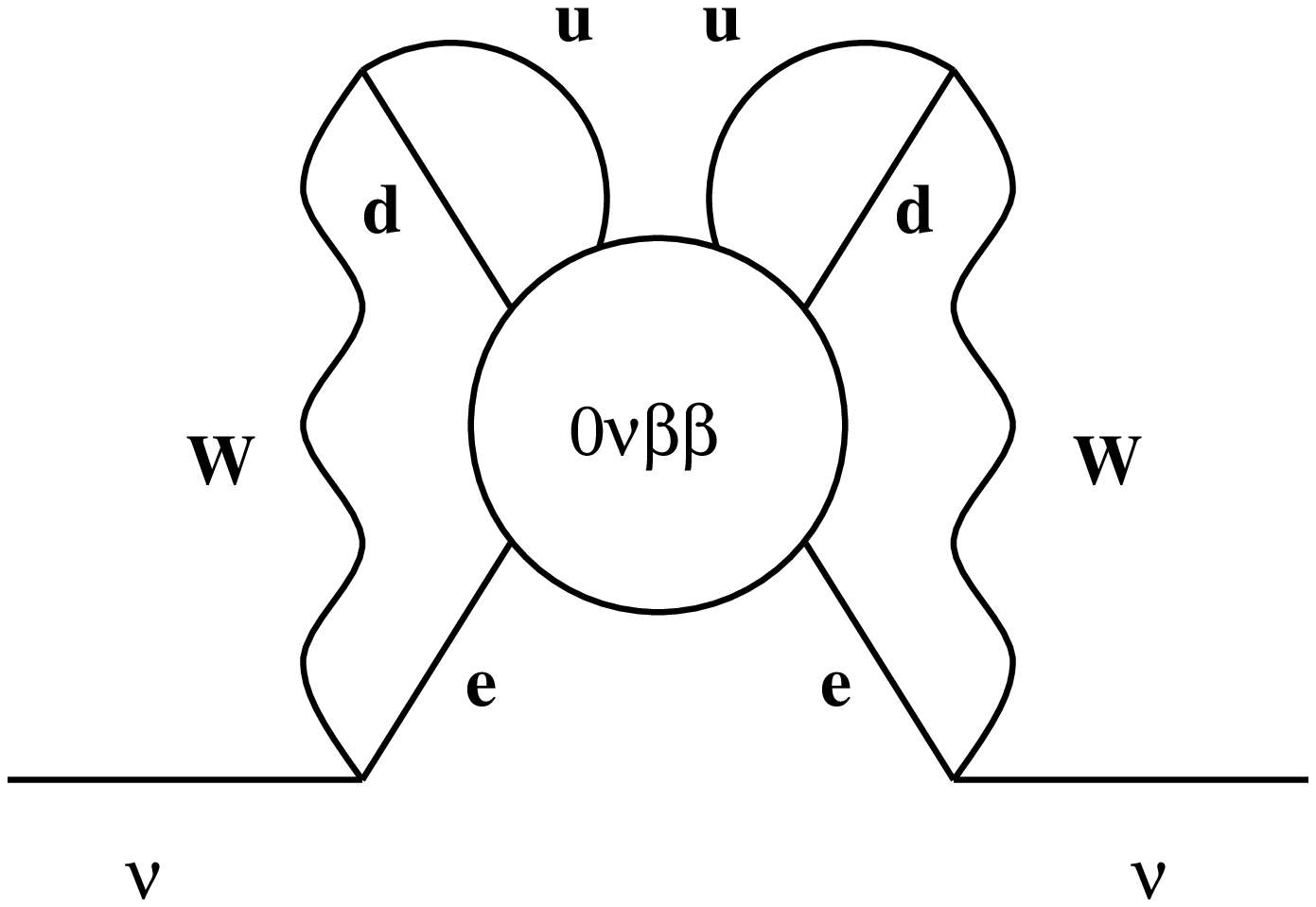}
\caption{Neutrino mass mechanism for \znbb
  (left), and black box theorem (right)~\cite{Schechter:1982bd}.}
 \label{fig:bbox}
\end{figure}

Neutrino oscillations can not distinguish Dirac from Majorana
neutrinos. In contrast, LNV processes, such as
\znbb~\cite{Schechter:1982bd} hold the key to the issue.
Indeed, in a gauge theory, \emph{irrespective of the mechanism that
  induces \znbb}, it implies a Majorana mass for at least one
neutrino~\cite{Schechter:1982bd}, as illustrated in Fig.
\ref{fig:bbox}.\\
\vskip .1cm

Such ``black-box'' theorem~\cite{Schechter:1982bd} holds in any
``natural'' gauge theory, though quantitative implications are very
model-dependent, for a recent discussion see \cite{Hirsch:2006yk}.
The detection of neutrinoless double beta decay remains a major
challenge~\cite{Avignone:2007fu}.

The observation of neutrino oscillations suggests that the exchange of
light Majorana neutrinos will induce \znbb through the so-called
\emph{mass-mechanism}. The corresponding amplitude is sensitive both
to the Majorana CP violation~\cite{schechter:1980gr}, and also to the
absolute scale of neutrino mass, neither of which can be probed in
oscillations. Together with high sensitivity beta decay
studies~\cite{Drexlin:2005zt}, and with cosmic microwave background
and large scale structure observations~\cite{Lesgourgues:2006nd},
neutrinoless double beta decay provides complementary information on
the absolute scale of neutrino mass.

Taking into account current neutrino oscillation
parameters~\cite{Maltoni:2004ei,Schwetz:2008er} and state-of-the-art
nuclear matrix elements~\cite{Faessler:2008xj} one can determine the
average mass parameter $\meff$ characterizing the neutrino exchange
contribution to \znbb, as in Fig. 42 of Ref.~\cite{Nunokawa:2007qh}.
Quasi-degenerate neutrino models~\cite{babu:2002dz,Hirsch:2003dr} give
the largest possible \znbb signal. In normal hierarchy models there is
in general no lower bound on $\meff$ as there can be a destructive
interference among the three neutrinos. In contrast, the inverted
neutrino mass hierarchy implies a generic ``lower'' bound for the
\znbb amplitude. Specific flavor models may, however, imply a lower
bound for \znbb even with normal hierarchy, as discussed
in~\cite{Hirsch:2009mx}~\cite{Hirsch:2005mc,Hirsch:2008mg,Hirsch:2008rp}.
The best current limit on $\meff$ comes from the Heidelberg-Moscow
experiment, for the current experimental status and perspectives, see
Ref.~\cite{Avignone:2007fu}, which should be compared with nuclear
theory~\cite{Faessler:2008xj}.

\vspace{.2cm}

{\em Acknowledgments:}

This work is supported by the Spanish grants FPA2008-00319/FPA and
PROMETEO/2009/091 and by European Union network UNILHC
(PITN-GA-2009-237920). I thank the organizers for hospitality in
Boston.  

\renewcommand{\baselinestretch}{1}

\bibliographystyle{aipproc}   

\end{document}